\documentclass[graybox,a4paper]{svmult}

\usepackage{mathptmx}
\usepackage{helvet}
\usepackage{courier}
\usepackage{type1cm}
\usepackage{makeidx}
\usepackage{graphicx}
\usepackage{multicol}
\usepackage[bottom]{footmisc}
\usepackage{overpic}
\usepackage{amssymb}

\newcommand{\ket}[1]{\left\vert#1\right\rangle}

\definecolor{stephen}{rgb}{0.8,0.1,0.5}

\makeindex

\begin{document}

\title*{Out-of-equilibrium physics in driven dissipative coupled resonator arrays}

\author{Changsuk Noh, Stephen R. Clark, Dieter Jaksch and Dimitris G. Angelakis}
\authorrunning{C. Noh,   S. R. Clark,  D. Jaksch, and D. G. Angelakis}

\institute{Changsuk Noh \at Centre for Quantum Technologies, National University of Singapore, 2 Science Drive 3, 117542, Singapore. \email{undefying@gmail.com}
\and Dimitris G. Angelakis \at Centre for Quantum Technologies, National University of Singapore, 2 Science Drive 3, 117542, Singapore \& School of Electrical and Computer Engineering, Technical University of Crete, Chania, Crete, 73100 Greece. \email{dimitris.angelakis@gmail.com}
\and Stephen R. Clark \at Clarendon Laboratory, University of Oxford, Parks Road, Oxford OX1 3PU, UK \&  Centre for Quantum Technologies, National University of Singapore, 2 Science Drive 3, 117542, Singapore \\
\and Dieter Jaksch \at Clarendon Laboratory, University of Oxford, Parks Road, Oxford OX1 3PU, UK  \& Centre for Quantum Technologies, National University of Singapore, 2 Science Drive 3, 117542, Singapore  \\}
\maketitle

\abstract{Coupled resonator arrays have been shown to exhibit interesting many-body physics including Mott  and Fractional Hall states of photons. One of the main differences between these photonic quantum simulators and their cold atoms counterparts is in the dissipative nature of their photonic excitations. The natural equilibrium state is where there are no photons left in the cavity.  Pumping the system with external drives is therefore necessary to compensate for the losses and realise non-trivial states. The external driving here can easily be tuned to be incoherent, coherent or fully quantum, opening the road for exploration of many body regimes beyond the reach of other approaches. In this chapter, we review some of the physics arising in driven dissipative coupled resonator arrays including photon fermionisation, crystallisation,  as well as photonic quantum Hall physics out of equilibrium. We start by briefly describing possible experimental candidates to realise coupled resonator arrays along with the two theoretical models that capture their physics, the Jaynes-Cummings-Hubbard and Bose-Hubbard Hamiltonians. A brief review of the analytical and sophisticated numerical methods required to tackle these systems is included.}

\section{Introduction}

One of the key problems in physics is to understand strongly correlated many-body systems. Fully solving a many-body Hamiltonian, analytically or numerically, is however a notoriously difficult problem. An increasingly plausible way around this issue is to use an experimentally accessible and well-controlled quantum system (e.g., cold atoms in optical lattices) to simulate the physics of another complex quantum system (e.g., electrons in a 2D lattice described by the Hubbard model) \cite{Feynman}. While cold atoms \cite{BlochNascimbene2012} and trapped ions \cite{BlattRoos2012} arose as strong candidates for quantum simulators \cite{CiracZoller2012, JohnsonJaksch2014} of interacting fermions/bosons and quantum spin systems, respectively, photonic systems based on linear optics and Circuit QED architectures were also recently proposed as alternative candidates with their own advantages \cite{Aspuru-GuzikWalther2012,HouckKoch2012}. One prominent example is the simulation of strongly interacting condensed matter models such as the Bose-Hubbard model \cite{Fisher} in coupled resonator arrays (CRAs) \cite{HouckKoch2012, TomadinFazio2010, CarusottoCiuti2013, SchmidtKoch2013}. 

The Bose-Hubbard (BH) model is a lattice model for a single bosonic species with nearest-neighbour hopping and an on-site interaction. In Ref.~\cite{Fisher}, it was shown that the competition between the two processes leads to a quantum phase transition between the superfluid and Mott-insulating phases, which has been demonstrated experimentally with ultracold atoms in an optical lattice \cite{JakschZoller1998,GreinerBloch2002}. To realise a photonic superfluid-Mott phase transition in a CRA one needs to build up a Bose-Hubbard-like Hamiltonian. The hopping between the lattice sites is naturally provided by an optical coupling between resonators, whereas to realise the on-site interaction, one has to induce some kind of nonlinearity in each resonator. For this purpose, an (artificial) atom can be coupled to each resonator, or for the case of solid-state cavities an intrinsic Kerr-nonlinearity may be considered. In the latter case, the dynamics of the CRA is governed by the BH model, whereas in the former case the dynamics depends on the type of the doped atom. When the atom is a two-level system, for which the coupling to the cavity mode is of the Jaynes-Cummings type, the resulting CRA is now known as the Jaynes-Cummings-Hubbard (JCH) system. 

Following an initial realisation that photons in CRAs undergo superfluid-Mott phase transition \cite{HartmannPlenio2006, Greentree, Angelakis}, the equilibrium phases of the JCH model has been extensively investigated \cite{Hartmann_review} (see also chapters by Schmidt \& Blatter and Tomadin \& Fazio). However, for photons it is more natural to consider a driven dissipative scenario, where photons are continuously pumped into the system to counteract unavoidable losses. This has led to the study of CRAs in non-equilibrium driven dissipative regimes, where strongly correlated many-body behaviour can be observed. The aim of this chapter is to discuss efforts in this direction. We discuss in detail a few topical examples, mostly on driven dissipative systems, but also on transient effects in dissipative arrays. We begin by providing a short survey of experimental platforms that could realise CRAs and then discuss generic models that describe these systems.

\subsubsection*{Candidates for realising CRAs}

{\it{Photonic crystals}} are defect semiconductors with holes periodically introduced to an otherwise uniform substrate.
\begin{figure}[h]
\centering
\includegraphics[width=7cm]{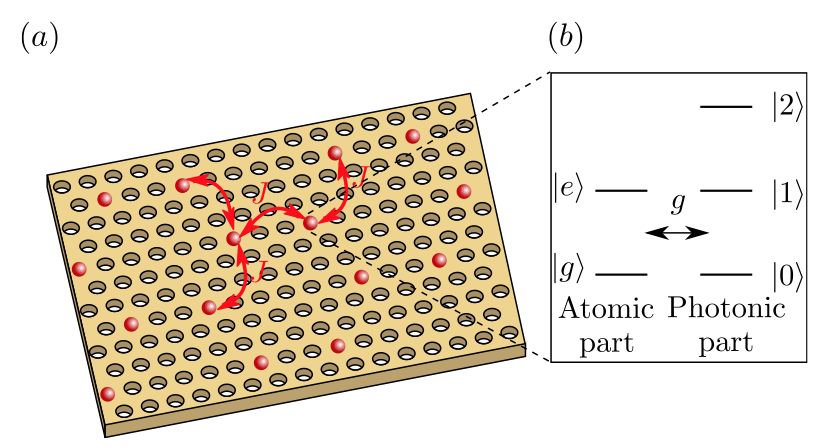}
\caption{Photonic crystal: (a) Periodic holes are introduced to modulate the refractive index of the material, giving rise to band gaps in the material. Defect cavities, represented by red circles, introduce isolated photonic states within the band gap. Photons can hop between neighbouring cavities, as represented by red arrows, due to modal overlap.
(b) When a quantum dot is grown inside the defect cavity, the interaction between the quantised electronic levels of the quantum dot and the cavity modes can be described by the Jaynes-Cummings model.
}
\label{fig:cra_photonic_crystal_cavities}
\end{figure}
This induces a spatial dependence of refractive index, and thus gives rise to photonic band gaps within the infra-red to visible range \cite{joannopoulos_johnson_winn}.
A cavity is introduced to the system by creating an isolated photonic mode within the band gap, for example by removing one or more holes, that are strongly confined to the photonic crystal plane. Figure \ref{fig:cra_photonic_crystal_cavities} illustrates such coupled photonic crystal cavities.
Photon-photon interactions can be induced by growing a semiconductor nanostructure called quantum dot in each cavity. Due to the tight confinement of electrons in a quantum dot, discrete energy levels form and when one of the transitions match the cavity frequency the resulting dynamics is well described by the Jaynes-Cummings Hamiltonian (see Sect.~\ref{section2} for a quick introduction to the Jaynes-Cummings model).
The advantages of photonics crystal cavities include low photon losses, and extremely small mode volumes on the order of a cubic wavelength $\lambda^3$.

In another promising semiconductor-based platform, a quantum well is sandwiched between two Bragg reflectors. The reflectors create an effective cavity, with a mode that interacts with the exciton in the quantum well. This creates the so-called exciton-polaritons which inherit nonlinearities from the electron-hole interaction. The interested reader is referred to chapters [Gerace] and [Kim] for more details on these devices. \\

\begin{figure}[h]
\centering
\includegraphics[width=8cm]{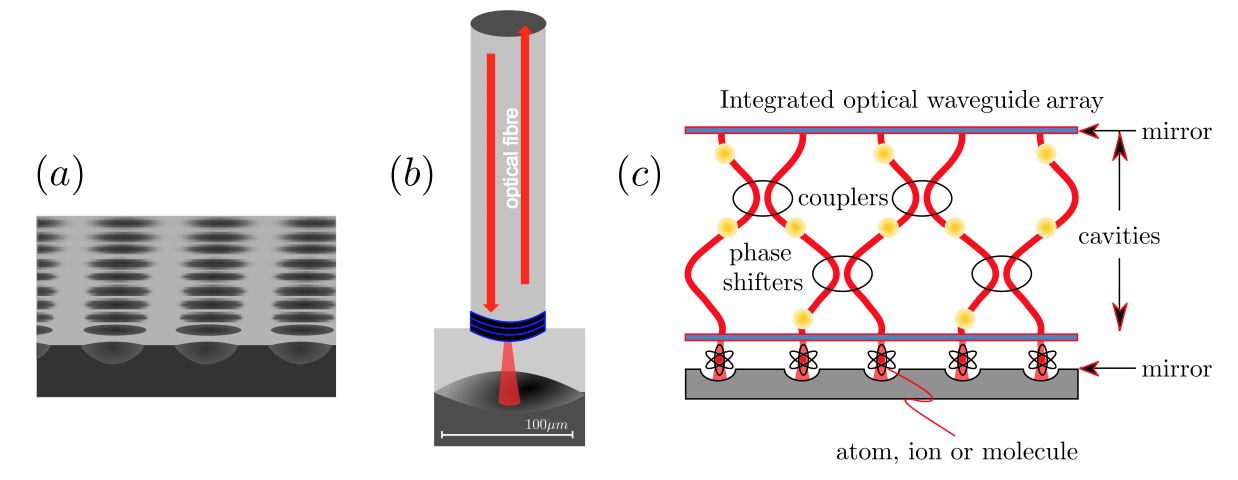}
\caption{
Integrated optical waveguide: 
(a) An array of periodic concave mirrors etched into Si can be fabricated with high-precision \cite{TrupkeKraft2005}, while (b) resonators may be formed by the addition of Bragg-inscribed fibres above them. 
(c) CRAs are realised by replacing the fibres with waveguides, inscribed with different refractive indices by a strong UV ``writing'' laser \cite{LepertSmith2013}. Photon tunneling between neighbouring waveguides/cavities is enhanced by optimising waveguide paths.
}
\label{fig:cra_integrated_optical_waveguides}
\end{figure}

\begin{figure}[h]
\centering
\includegraphics[width=7cm]{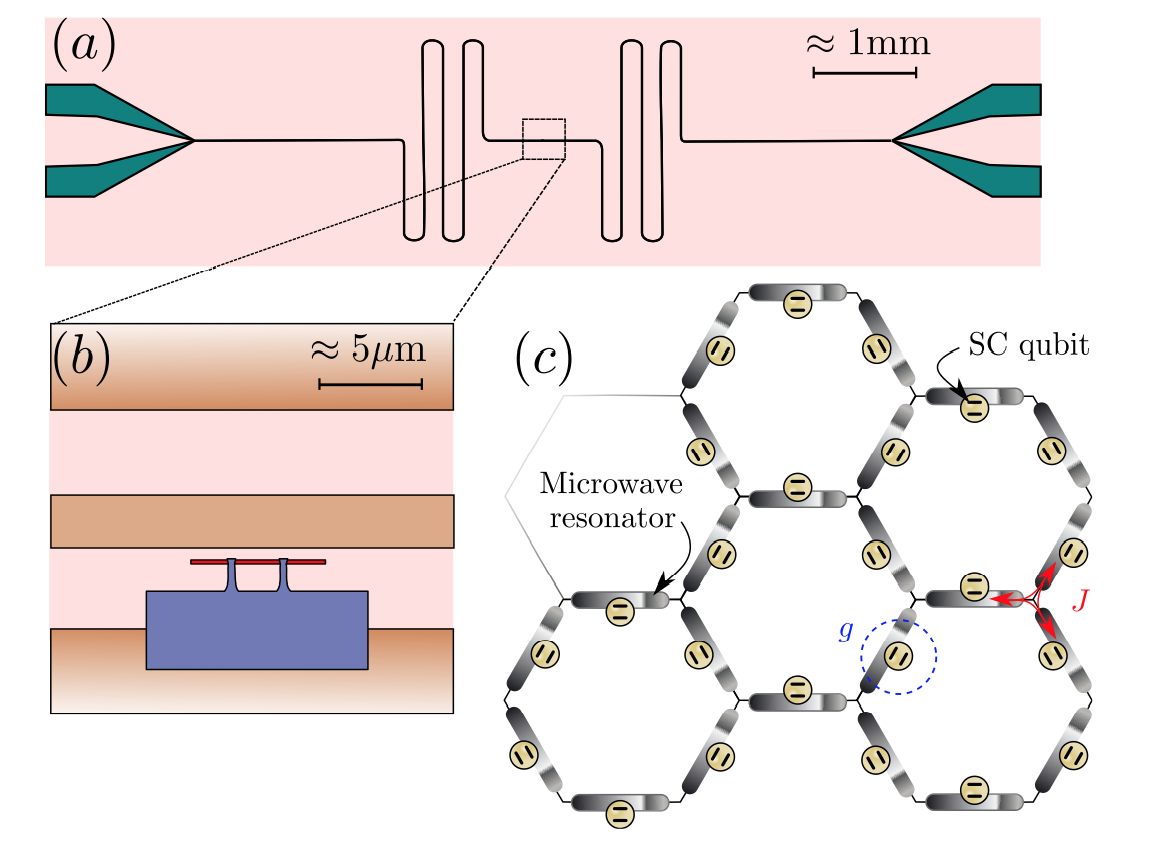}
\caption{
Superconducting circuit QED:
(a) A superconducting coplanar waveguide resonator fabricated on a silicon chip using optical lithography. Green tapering structures fanning out to the edges of the chip are input and output feed lines. The center conductor is separated from the lateral ground planes by a gap as shown in the zoom.
(b) Effective 2-level system, a superconducting qubit, coupled to the coplanar waveguide resonator. A superconducting island connected by a pair of Josephson junctions to the superconducting reservoir is placed in the gap between the central conductor and the bottom ground plane. 
(c) Many stripline resonators fabricated into a JCH system.
}
\label{fig:cra_superconducting_cqed}
\end{figure}

{\it{Integrated waveguide arrays}} are another interesting candidate. In this set-up, the formation of resonators involve precise etching of concave mirror arrays into silicon \cite{TrupkeKraft2005} with Bragg-inscribed optical fibres placed above them (see Fig. \ref{fig:cra_integrated_optical_waveguides}(a) and (b)). CRAs can be realised by introducing laser-written waveguides in places of optical fibres \cite{LepertSmith2013}, such that the photon tunneling is allowed by evanescent coupling between the waveguides (see Fig. \ref{fig:cra_integrated_optical_waveguides}(c)). Phase shifters can then be used to tune the coupling strength, while the nonlinearity can be introduced by placing an atom on each concave mirror cavity. 
\\
{\it Superconducting circuits}, our last candidate, is the most promising platform for realising CRAs \cite{HouckKoch2012, blais_huang_wallraff, astafiev_zagoskin_abdumalikov, frunzio_wallraff_schuster,  devoret_girvin_schoelkopf}. They enjoy low photon loss rates due to superconductivity and good scalability due to highly-accurate fabrication procedures available. Various circuit designs exist to create an artificial two-level atom, all of which however share the common feature of exploiting the intrinsic nonlinearity of one or more Josephson junctions. Such an artificial atom can be coupled to superconducting transmission line resonators to mimic the Jaynes-Cummings interaction in cavity QED, as shown in Fig.~\ref{fig:cra_superconducting_cqed} \cite{wallraff_schuster_blais}. The resulting system is often called the circuit QED (cQED) system in analogy to cavity-QED. A comparison between the various realisations of CRAs using currently available parameters are summarised in Table \ref{table1}. Note that a large cooperativity (the ratio between the interaction strength $g$ and the loss rate $\gamma$) is required in most proposals for equilibrium simulations.
\begin{table}[ht]
\caption{Comparison of typical CRA parameters for different architectures. Note that the values quoted for integrated optics platform are not taken from actual experiments, but are realistic parameters predicted in ref.~\cite{LepertHinds2011}.}
\centering
\begin{tabular}{l c c c c}
\hline\hline
Parameter & Symbol\,\,  & Photonic crystals \,\,  & Integrated optics \,\, & Superconducting \\ [0.5ex] \hline
Resonance frequency & $\omega_c/2\pi$ & 325 THz & 380 THz & 10 GHz \\
JC parameter & $g/2\pi$, $g/\omega_c$ & 20 GHz  & 33 MHz & 200 MHz \\
Cavity decay rate & $\gamma/2\pi$ & 1 GHz & 10 MHz & 100 kHz \\
Atom decay rate & $\kappa/2\pi$ & 8 GHz & 10 MHz & 2 kHz \\
Cooperativity & $g/\gamma\kappa$ & 2.5 & 10 & $\gg 1$ \\
Resonator coupling & $J/2\pi$ & 100 GHz & 2 GHz & 10 MHz \\ [1ex]
\hline
\end{tabular}
\label{table1}
\end{table}

\section{Modelling CRAs}
\label{section2}

There are two models often used to describe CRAs, depending on whether the nonlinearity is assumed to be intrinsic to the resonator, or is explicitly introduced by coupling to each resonator a two-level atom. When losses are neglected, the former case is described by the BH model (from here on, we assume $\hbar = 1$)
 \begin{equation}
\hat{H}_\mathrm{BH} = -J \sum_{\langle i,j\rangle} \left(\hat{a}_i^\dag \hat{a}_j + {\rm h.c.} \right) + \frac{U}{2} \sum_i  \hat{a}_i^\dag \hat{a}_i^\dagger \hat{a}_i\hat{a}_i + \omega_c \sum_i \hat{a}_i^\dagger \hat{a}_i,
\label{eq:bh}
\end{equation}
whereas the latter is described by the JCH model:
\begin{eqnarray}
\hat{H}_\mathrm{JCH} = -J \sum_{\langle i,j \rangle} \left(\hat{a}^\dag_i \hat{a}_j + \mathrm{h.c.} \right)   + g \sum_i \left(\hat{a}^\dag_i \hat{\sigma}^-_i + \mathrm{h.c.} \right) \nonumber \\ + \omega_c \sum_i \left(\hat{a}^\dag_i \hat{a}_i + \hat{\sigma}^+_i \hat{\sigma}^-_i \right) - \Delta \sum_i \hat{\sigma}^+_i \hat{\sigma}^-_i.
\label{eq:jch_hamiltonian}
\end{eqnarray}
Here $\hat{a}$ is the photon annihilation operator, $\hat{\sigma}_-$ is the atomic lowering operator, $J$ is the photon hopping rate between neighbouring cavities, $U$ is the intrinsic nonlinearity of the resonator, $g$ is the coupling strength between the resonator and the two-level system, $\omega_c$ is the natural frequency of the resonator, and $\Delta$ is the detuning between the resonator and the two-level system. When $J=0$, $\hat{H}_{JCH}$ becomes the Jaynes-Cummings model, which describes the physics of a two-level atom coupled to a resonator mode when the coupling strength $g$ is much smaller than the atomic and resonator frequencies \cite{WallsMilburn}. The Kerr-nonlinearity in the BH model also comes from some sort of interaction with matter, but in the regime where only the overall effects of the material on the resonator mode are considered. These effects are usually weak, but can be enhanced in some systems as reviewed in \cite{CarusottoCiuti2013}.

With losses included, the Hamiltonian description is no longer sufficient. However, it is possible to model such open systems by coupling the system of interest to a bath of harmonic oscillators. Assuming weak system-bath coupling and a few general properties of the bath, one can obtain an equation of motion for the reduced density matrix of the system alone \cite{WallsMilburn}. The resulting quantum master equation (QME) can be written as 
\begin{equation}
d\rho/dt = -i \left[\hat{H}_\alpha , \rho \right] + \sum_j  \frac{\gamma_j}{2} \left(2\hat{a}_j \rho \hat{a}_j^\dag - \hat{a}_j^\dag \hat{a}_j \rho - \rho \hat{a}_j^\dag \hat{a}_j  \right),
\label{mastereqn}
\end{equation}
where $\rho$ is the density operator and $\alpha \in \{ {\rm JCH}, {\rm BH}\}$. The second term on the right hand side is a Lindblad noise term we will call $\mathcal{L}_{\rm loss}$, which accounts for the dissipation in the cavities where $\gamma_j$ is the loss rate of cavity $j$.
For the JCH systems, an additional source of dissipation, due to decay of the excited atomic state, can also be taken into account by adding another Lindblad noise term
\begin{equation}
\mathcal{L}_{\rm decay}\{\rho\} =  \sum_j  \frac{\kappa_j}{2} \left(2\hat{\sigma}^-_j \rho \hat{\sigma}^+_j - \hat{\sigma}^+_j \hat{\sigma}^-_j \rho - \rho \hat{\sigma}^+_j \hat{\sigma}^-_j \right),
\label{mastereqn}
\end{equation}
where $\kappa_j$ is the decay rate for the $j$th two-level atom. To compensate for the losses, experimentalists usually drive one or more resonators with lasers. Such driving can be modelled by adding the term
\begin{equation}
\hat{H}_{\rm drive} =  \sum_j \Omega_j(t)\hat{a}_j^\dagger +\Omega_j^*(t)\hat{a}_j,
\end{equation}
to the Hamiltonian, where $\Omega_j(t) = \Omega_je^{-i\omega_L t}$ is the Rabi frequency of the driving laser. Incoherent driving can also be added in the form of a Lindblad noise operator\footnote{One way to see this is to derive the rate equation for the diagonal density operator elements from the master equation.}. For example, the cavity array driven by a thermal reservoir field with average photon number $\bar{n}$ is described by 
\begin{eqnarray}
d\rho/dt &=& -i \left[\hat{H}_\alpha , \rho \right] + (\bar{n} +1)\sum_j  \frac{\gamma_j}{2} \left(2\hat{a}_j \rho \hat{a}_j^\dag - \hat{a}_j^\dag \hat{a}_j \rho - \rho \hat{a}_j^\dag \hat{a}_j  \right) \nonumber \\ &&+ \bar{n}\sum_j  \frac{\gamma_j}{2} \left(2\hat{a}_j^\dag \rho \hat{a}_j - \hat{a}_j \hat{a}_j^\dag \rho - \rho \hat{a}_j \hat{a}_j^\dag  \right).
\end{eqnarray}

\section{Computing the properties of CRAs}
Given the QME description of a CRA, in most cases we will concentrate on the properties of the non-equilibrium steady state (NESS), found by solving $d\rho/dt = 0$, in analogy to the ground state properties of the equilibrium systems. We will also be interested in the transient open system dynamics given the system is prepared in some initial state. For either task we are faced with integrating the QME for a BH or JCH system -- a technically challenging problem due to the exponential growth in the Hilbert space dimension with the number of sites (e.g. resonators or resonator + atoms) considered. Developing numerical algorithms to aid in studying driven dissipative strongly-correlated many-body quantum systems is one of the most difficult problems in computational physics. However, the last decade has seen impressive progress in the development of computational representations of the state of quantum lattice systems as tensor networks \cite{verstraete_murg_cirac,cirac_verstraete,tnt_library}, in particular in 1D via so-called matrix product states (MPS) \cite{perez_garcia,schollwock_ann_phys}. 

The key feature of MPS is that they provide a compact and efficient ansatz for describing weakly-entangled pure quantum states well suited to open chains of sites \cite{schollwock_ann_phys}. This has been used with much success to describe the low-lying excited states and zero-temperature unitary dynamics of cold-atom systems, e.g. the $-i\left[ H'_\alpha , \rho \right]$ term in the QME \cite{white_a,white_b,vidal_a,vidal_b,schollwock_rmp}. These methods have been readily adapted \cite{zwolak_vidal,verstraete_garcia_ripoll,daley} to describe dissipative systems with Lindblad noise terms which are local incoherent processes, like $\mathcal{L}_{\rm loss}$ and $\mathcal{L}_{\rm decay}$ relevant to CRA. This is done by unravelling the QME into quantum trajectories where many evolutions of pure states, described by an MPS, are made and stochastically interrupted by quantum jumps \cite{gardiner_parkins_zoller,dum_gardiner,dalibard_castin_molmer,carmichael,plenio_knight}. By averaging over many trajectories then recovers the properties of the full density matrix $\rho$ \cite{breuer_petruccione}. In the Appendix we describe in more detail this very powerful method.

\section{Non-equilibrium many-body phases of photons in CRAs}

In this section, we will review in detail a selection of many-body phenomena arising in non-equilibrium scenarios. To set the background, we start with a comparison between the physics of JC and Kerr-nonlinear resonator, followed by a description of strong photon-bunching phenomenon in a driven dissipative two-site (dimer) case. Next, a transient phenomenon called localisation-delocalisation of photons in a JCH dimer is presented, which shows a nice example of how CRAs are implemented in a circuit-QED platform. Continuing our review with driven dissipative phenomena, we discuss fermionisation and crystallisation of photons in 1D CRAs as well as the NESS phases and quantum Hall-like physics in 2D arrays. Last subsection provides a brief survey of other interesting driven dissipative studies that we do not have enough space to cover in detail.

\subsection{Jaynes-Cummings vs Kerr-nonlinear resonator}

\begin{figure}[ht]
\centering
\includegraphics[width=9cm]{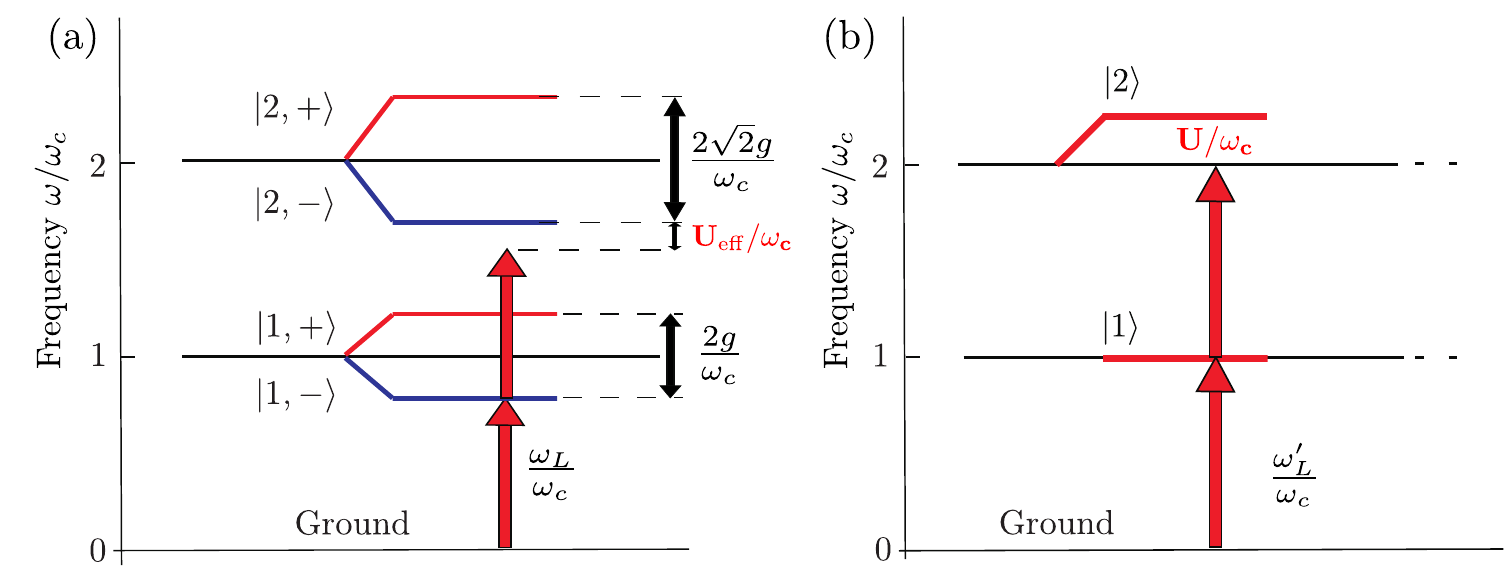}
\caption{Low-lying eigenstates of the (a) Jaynes-Cumming model and the (b) Bose-Hubbard model. Effective photon-photon interaction is provided by the anharmonicity in the spectrum of a given polariton branch. Taken from Grujic et al.~\cite{GrujicAngelakis2012}.}
\label{tomfig1}
\end{figure}

In the limit of zero atom-photon detuning, the eigenstates of the Jaynes-Cummings Hamiltonian, i.e., the $J=0$ and $\Delta=0$ case of (\ref{eq:jch_hamiltonian}), are the polariton (dressed photon-atom) states $|m,\pm\rangle = \frac{1}{\sqrt{2}} \left( |G,m\rangle \pm |E, m-1\rangle \right)$ having eigenenergies $\mathcal{E}_{m\pm} = \omega_c m \pm g \sqrt{m}$ (see Fig.~\ref{tomfig1}(a)). 
Here, $G$ and $E$ refer to the ground and excited states of the atom, while $m$ is the number of photons in the resonator.
Imagine driving the lowest `$-$ polariton' branch $|1,-\rangle$ by setting $\omega_L = \omega_c - g$. Because of the nonlinearity induced by the atom, the transition to the excited state $|2,-\rangle$ is detuned by the amount $\mathcal{E}_{2-} - 2\mathcal{E}_{1-} = g(2-\sqrt{2})$. Compare this to the single-site BH system--a Kerr-nonlinear resonator--where the detuning is simply $U$ as depicted in Fig.~\ref{tomfig1}(b). When photons are restricted to the lowest manifolds, we may therefore take the effective on-site interaction strength of the JC system as $U_{eff} = g(2-\sqrt{2})$ for zero atom-photon detuning. For non-zero detuning the expression becomes
\begin{equation}
\frac{U_{eff}}{g} = \frac{\Delta}{2g} + 2\sqrt{\left( \frac{\Delta}{2g}\right)^2 + 1} - \sqrt{\left( \frac{\Delta}{2g}\right)^2 + 2}.
\label{ueff}
\end{equation}

\begin{figure}[ht]
\centering
\includegraphics[width=4cm]{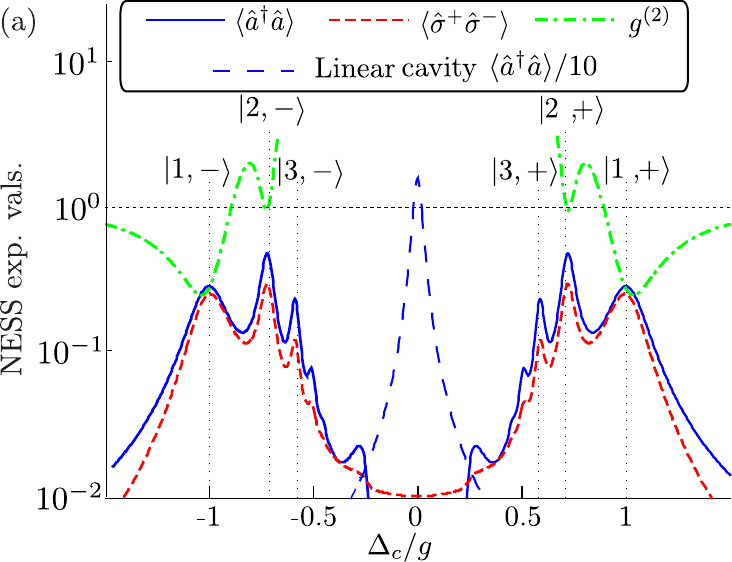}
\includegraphics[width=4cm]{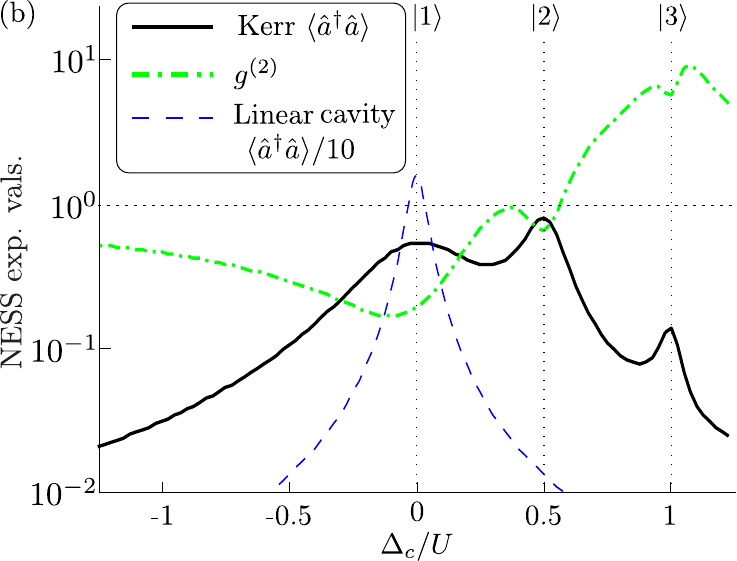}

\vspace{0.1cm}
\includegraphics[width=4cm]{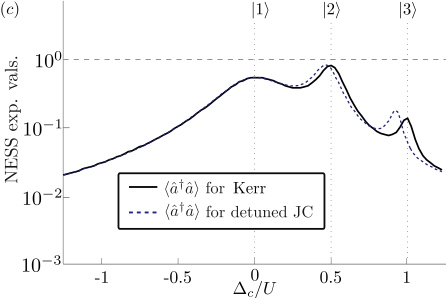}
\caption{
Steady state values of photonic/atomic population (blue solid/red dashed curves) and the second-order intensity correlation function $g^{(2)}$ (green dot-dashed curve) as functions of the driving laser-cavity detuning. (a) Jaynes-Cummings resonator with zero atom-photon detuning. The Lorentzian profile of the empty driven dissipative resonator is shown for comparison (blue dashed curve). (b) Kerr-nonlinear resonator. Vertical dashed lines mark expected resonances in each model, corresponding to the labeled states. Parameters are: $\Omega/\gamma = 2$, $g/\gamma = 20$ (JC), and $U/\gamma \approx 11.7$ (Kerr, as calculated from Eq.~(\ref{ueff})). (c) Comparison of the steady-state photon numbers for a Kerr resonator with parameters as used in (b) and a JC resonator with $\Delta/g = -10$ and $g/\gamma \approx 1.6 \times 10^4$. Taken from Grujic et al.~\cite{GrujicAngelakis2012}.}
\label{fig:singlecavity}
\end{figure}

The `repulsion' between photons induced by the nonlinearity can be quantified by the (equal-time) second-order intensity correlation function
\begin{equation}
g^{(2)}  = \langle \hat{a}^\dagger \hat{a}^\dagger \hat{a} \hat{a} \rangle/\langle \hat{a}^\dagger \hat{a} \rangle^2,
\end{equation}
as shown in Fig.~\ref{fig:singlecavity}.
Note the dips in $g^{(2)}$ when the laser field (with laser-cavity detuning $\Delta_c = \omega_L - \omega_c$) is resonant with the eigenstates of each model. In particular, the dip is most prominent and falls below 1 when the lowest excited state is driven (the situation depicted in Fig.~\ref{tomfig1}), signifying that the presence of a photon in a resonator prevents another one from entering the it. This phenomenon is called photon-blockade or photon anti-bunching. Stronger the repulsion, stronger the blockade or anti-bunching effect, i.e., $g\rightarrow 0$.

Despite this similarity between the JC and Kerr resonators, there are qualitative differences which become important when modelling CRAs as first investigated by Grujic et al.~\cite{GrujicAngelakis2012}. Let we look at the single resonator case for simplicity.
First, broad similarities between the two models are apparent upon comparing the left wing of Fig.~\ref{fig:singlecavity}(a) (corresponding to the lower ($-$) polariton branch) to Fig.~\ref{fig:singlecavity}(b): there are peaks at the eigen-modes of the resonators accompanied by the dips in $g^{(2)}$. However, the JC model displays richer structure, as indicated by the presence of two symmetric wings (corresponding to the two species of polaritons) and shift in the lowest excitation state with respect to the linear case (due to the atom-photon interaction). 

It is then natural to ask if the two models can be made equivalent at all, so that the physics of the JC resonator can be described by an effective Kerr model. Indeed this is possible as shown in Fig.~\ref{fig:singlecavity}(c). To achieve good agreement, one needs to take the photonic limit of the JC model in which the photonic component of the polaritons are dominant. For the $-$ polaritons this requires setting a large negative detuning $\Delta \ll -g$. However, this means reduced effective atom-photon interaction and $g$ must be increased in order to compensate for it . In Fig.~\ref{fig:singlecavity}(c), $g/\gamma \approx 1.6 \times 10^4$ was required to achieve good agreement for $\Delta/g = -10$. 
%


The above example shows that the JC and Kerr resonators are qualitatively different systems unless care is taken to ensure that they are in the right parameter regimes. The same conclusion holds for an array of resonators, meaning that, within realistic regimes of interactions, a phenomenon found in one type of array is not guaranteed to be reproduced in another. This provides flexibility in choosing CRAs and enriches physics that can be found in them. Later in this section, we will see further examples that highlight the similarities and differences between the two systems.


 

%
\subsection{Photon super-bunching}
\label{sect:superbunching}

Let us start with the simplest system of 2 coupled Kerr-nonlinear cavities and investigate the effects of an interplay between the hopping and on-site interaction. Here we focus on the small-sized equivalent of the `Mott' state, meaning that the driving field will be set on two-photon resonance with the lowest-energy eigenstate in the two-photon manifold $|E^{(2)}_-\rangle \propto |2,0\rangle + |0,2\rangle +(U+\sqrt{U^2 + 16J^2})/2\sqrt{2}J|1,1\rangle$. That is, $\omega_L = \omega_c + (U-\sqrt{U^2+16J^2})/4$. Note that this state converges to the `Mott' state $|1,1\rangle$ as $U\rightarrow \infty$. 

\begin{figure}
\centering
\includegraphics[width=8cm]{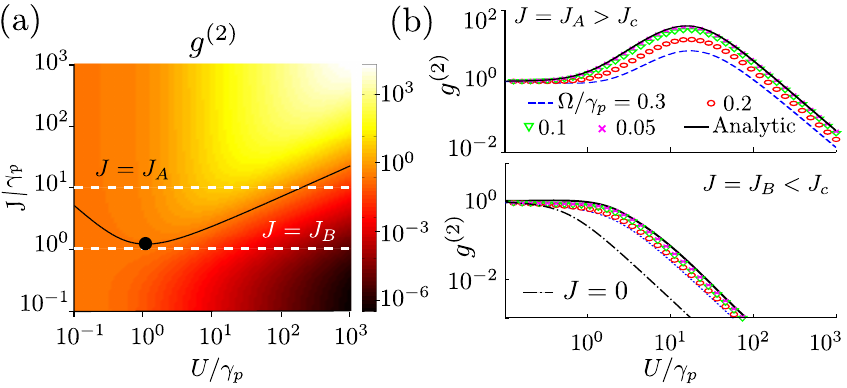}
\caption{
Photon super-bunching due to photon tunneling in the presence of the Kerr nonlinearity.
(a) Emitted photon statistics $g^{(2)}$ as a function of the photon tunneling strength $J$ and Kerr nonlinearity $U$.
Above some critical threshold $J_c$ indicated by the black curve, photon emission is bunched ($g^{(2)} > 1$).
Otherwise, photon emission is anti-bunched.
(b) Emitted photon statistics $g^{(2)}$ as a function of Kerr nonlinearity $U$ above and below the critical tunneling strength $J_c$. Super-bunching can be observed for tunneling strengths larger than $J_c$. We note that the definition of $U$ used in the figure is two times larger than the definition used in Eq.~(\ref{eq:bh}). Taken from Grujic et al.~\cite{GrujicAngelakis2013}.}
\label{fig:superbunching_bh}
\end{figure}

Studying this set-up, Grujic and coworkers have discovered that strong photon bunching  ($g^{(2)} \gg 1$) can be observed even when $U$ is significantly stronger than $J$ \cite{GrujicAngelakis2013}. 
This is depicted in Fig.~\ref{fig:superbunching_bh}, which plots the local $g^{(2)}$ of the emitted photons as a function of $J$ and $U$. Note that the diagram is broadly divided into three regions, defined by the Poissonian ($g^{(2)}\approx 1$), anti-bunched ($g^{(2)} < 1$), and bunched ($g^{(2)} > 1$) statistics of the emitted photons. The crossover points from anti-bunching to bunching are depicted by the black line, showing that there is a critical coupling strength, $J_c$, below which no bunching can be observed. This is further illustrated in Fig.~\ref{fig:superbunching_bh}(b), where the auto-correlations along the two white lines in (a) are drawn for various values of the driving strength. 
In both the linear ($U \ll \gamma$) and the hardcore ($U \gg \gamma$) limit, expected behaviours are observed: in the linear regime, the steady state is a coherent state, inheriting the Poissonian statistics from the driving laser, while in the hardcore regime, no more than a single photon per resonator can be injected, resulting in completely anti-bunched light. The behaviour of photons in between these extreme limits, however, depends strongly on the actual values of $U$ and $J$. For $J > J_c$, there can be strong bunching ($g^{(2)} \gg 1$) even with significant repulsion between photons $U>J$. Similar behaviour was found in the JCH dimer also \cite{GrujicAngelakis2013}.

The origin of strong bunching can be traced to the relative enhancement of the two-photon sector in the NESS, relative to the single-photon sector, as the nonlinearity is increased. In the linear case the NESS is a coherent state, meaning that the probability of having two-photons in the second cavity is equal to the square of the probability to have a single photon, giving rise to the observed coherence. Initially with increasing $U/J$, this distribution is modified such that the probability of having two photons is enhanced, i.e., it is larger than the square of the probability of having a single photon, because the driving field is detuned from the single-particle manifold. This explains the initial rise in $g^{(2)}$. However, this trend cannot continue indefinitely because the $|2,0\rangle + |0,2\rangle$ portion in the driven two-photon eigenstate decreases with with increasing $U/J$. At some point the $|1,1\rangle$ part of the two-photon eigenstate starts to dominate and the system exhibits antibunching. The interplay between the enhanced total two-photon sector and the suppressed local two-photon sector with increase in $U/J$ gives rise to the observed behaviour in $g^{(2)}$.

 
%
\subsection{Localisation-delocalisation transition of photons}
\label{sect:frozen}

Instead of continuously driving the system, let us now prepare the system in a certain state and observe its dynamics. The steady state of the system will be the trivial vacuum state, but we can still observe interesting physics in the transient dynamics. In particular, strong atom-photon interaction gives rise to a dramatic transition from a localised phase to a delocalised phase as first discovered by Schmidt et al.~\cite{SchmidtTureci2010}. In this set-up, photons are initially localised within a single site of a dimer. Without atom-photon coupling, the initially localised photons hop into the unoccupied cavity. However, as the atom-photon interaction is increased, there is a sharp transition to the localised regime where photons are trapped in the initial cavity. Similar delocalisation-localisation transition, or photon {\it freezing}, was first discovered in the BH-type systems and is known as self-trapping \cite{Jensen, Eilbeck, Smerzi, Albiez, Levy, Sarchi}. The result has been generalised to larger arrays by Schetakis et al.~\cite{SchetakisAngelakis2013}.

\begin{figure}[ht]
\centering
\includegraphics[width=5cm]{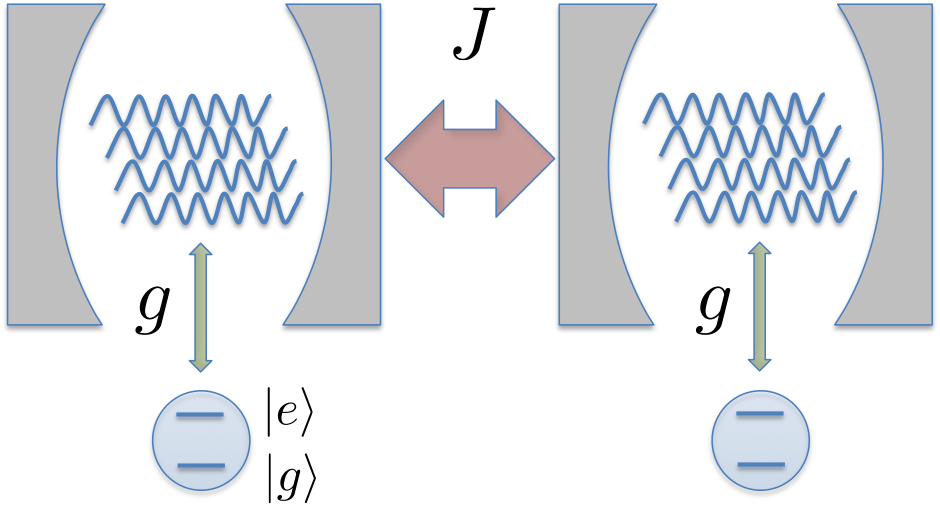}
\caption{
Schematic representation of the JCH system exhibiting `freezing' of photons. The left cavity of a dimer is initially pumped to contain 4 photons per resonator. The tunneling ratio is $J$ and the atom-photon coupling strength is $g$. 
}
\label{fig:freezing_scheme}
\end{figure}

The system is represented schematically in Fig.~\ref{fig:freezing_scheme}. There are $M$ resonators with the left-side resonators initially occupied by $N_0$ photons per resonator with the atoms in the ground state. The transition between localised and delocalised phases is nicely captured by the photon imbalance defined as
\begin{equation}
Z(t) = \frac{\sum_{j =1}^{M/2} \langle \hat{a}^\dag_j (t) \hat{a}_j (t) \rangle - \sum_{j =M/2+1}^{M}   \langle \hat{a}^\dag_j (t) \hat{a}_j (t) \rangle}{\sum_{j =1}^{M} \langle \hat{a}^\dag_j (t) \hat{a}_j (t) \rangle},
\end{equation}
or rather the time-averaged version of it, which we will denote as $\bar{Z}$. 
Using this definition, the localised regime is characterised by $|\bar{Z}| \approx 1$ and the delocalised regime by $\bar{Z} \approx 0$.

\begin{figure}[ht]
\centering
\includegraphics[width=4cm]{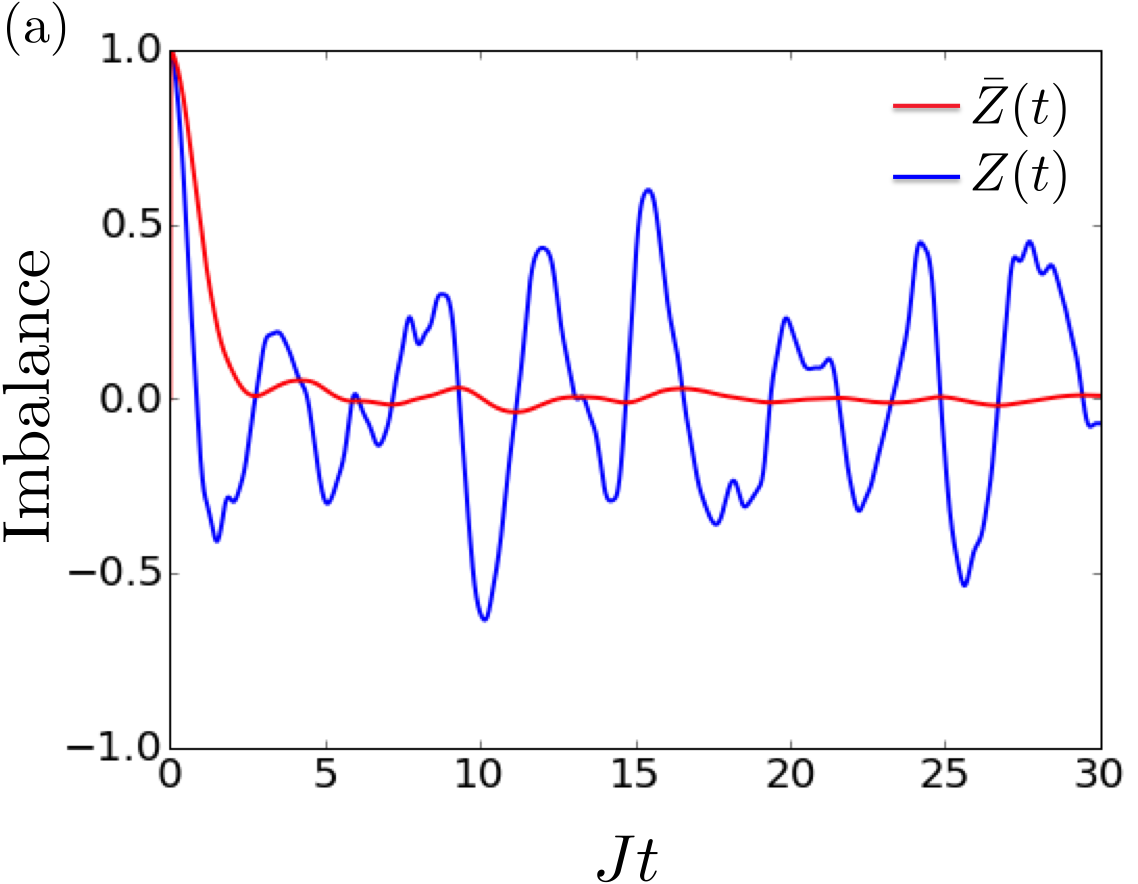} \hspace{0.2cm}
\includegraphics[width=4cm]{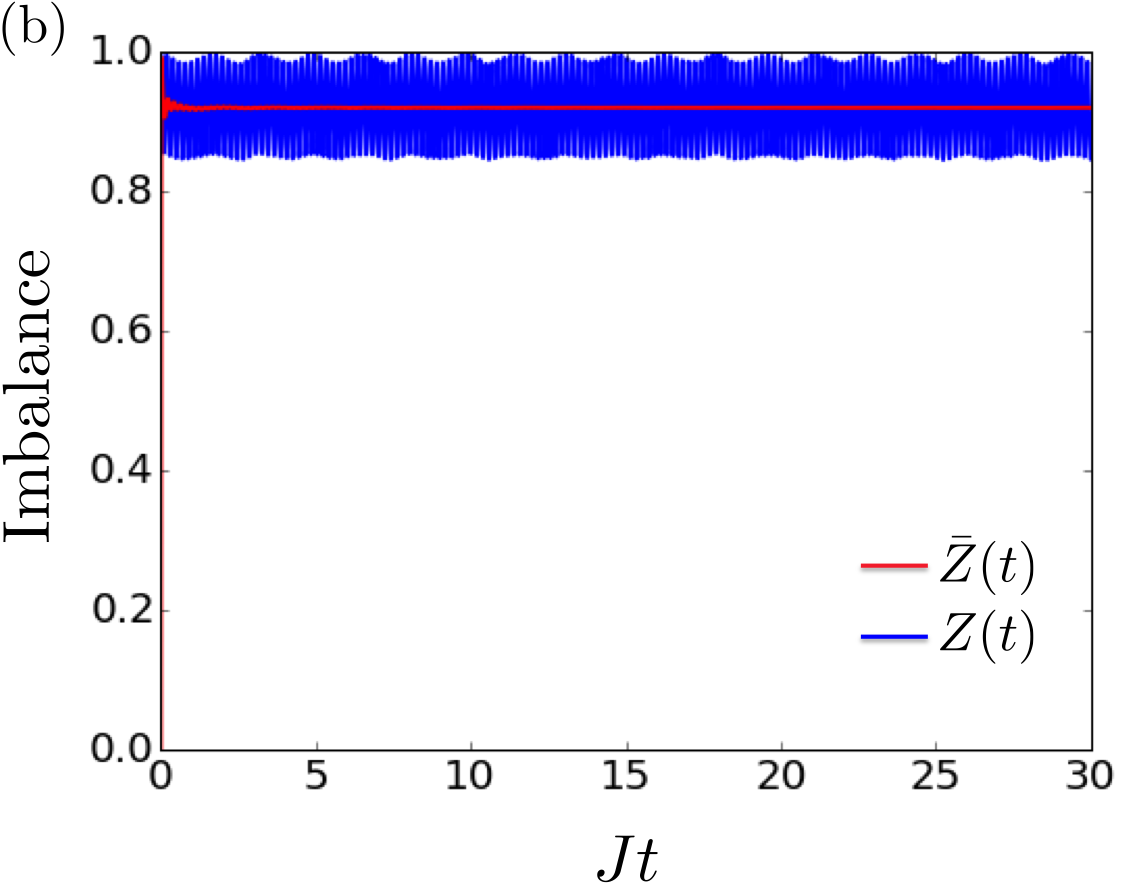}
\caption{
Photon number imbalance (blue curves) and the time-averaged imbalance (red curves) for the lossless dimer with 7 initial photons. (a) Below the critical coupling strength, $g=0.3g_c^{cl}$, showing delocalised dynamics. (b) Above the critical coupling strength, $g=2.0g_c^{cl}$, showing localised dynamics. 
}
\label{fig:photonnumber}
\end{figure}

Figure~\ref{fig:photonnumber} provides a succinct account of the delocalised (left hand side) and localised (right hand side) dynamics for the lossless dimer with 7 initial photons. It shows the imbalance $Z$ and the time average of it ($\bar{Z}(t)$) for two values of atom-photon coupling strength: One well below the critical coupling strength, $g=0.3g_c^{cl}$ (left hand side), and the other is well above it, $g=2.0g_c^{cl}$ (right hand side), where $g_c^{cl}$ is the critical coupling strength which in the semiclassical approximation can be derived as $g_c^{cl} \approx 2.8\sqrt{N_0}J$ \cite{SchmidtTureci2010}. 
Below the critical coupling strength, photons are free to travel back and forth between the cavities and the average imbalance $\bar{Z}$ goes to zero with time. Above the critical strength, however, photons are self-trapped and cannot move to the right half of the array (rapid oscillations in the photon number is due to an energy exchange with the atom). This self-trapping behaviour stems from the fact that, for the symmetric system, the two states with opposite imbalance 1 and -1 are approximate eigenstates of the system. For the dissipationless dimer, the time scale of the oscillation between the states of opposite imbalance can be found within the degenerate perturbation theory as $1/(c_{N_0} J(J/g)^{N_0-1})$ with $c_{N_0}$ a constant that depends on $N_0$ \cite{SchmidtTureci2010}. In the strongly nonlinear regime, the time scale diverges with increasing $N_0$, i.e., the photons are effectively trapped in the initial site.

Detailed calculation shows that the localisation transition is shifted to smaller $g$ values and the transition is smoothened in the quantum case, i.e. $g^{qu}_c < g^{cl}_c$. With dissipation, the imbalance goes to zero trivially as the photons leave the system, but within the decay time, the localisation behaviour can be clearly observed. Interestingly, the dissipation can help one to achieve the transition from the delocalised regime to the localised regime. This follows from the fact that the critical coupling strength $g_c^{qu}$ increases with the number of photons in the system (as described by the semiclassical approximate formula $g_c^{cl} \propto \sqrt{N_0}J$). Consider an experimental situation in which $g$ is fixed and less than $g_c^{qu}$. As the photons leave the system, the total photon number decreases and consequently $g_c^{qu}$ decreases along with it. This means that $g_c^{qu}$ will eventually dip below $g$, at which point the system enters the localised regime. 

\begin{figure}[ht]
\centering
\includegraphics[width=5cm]{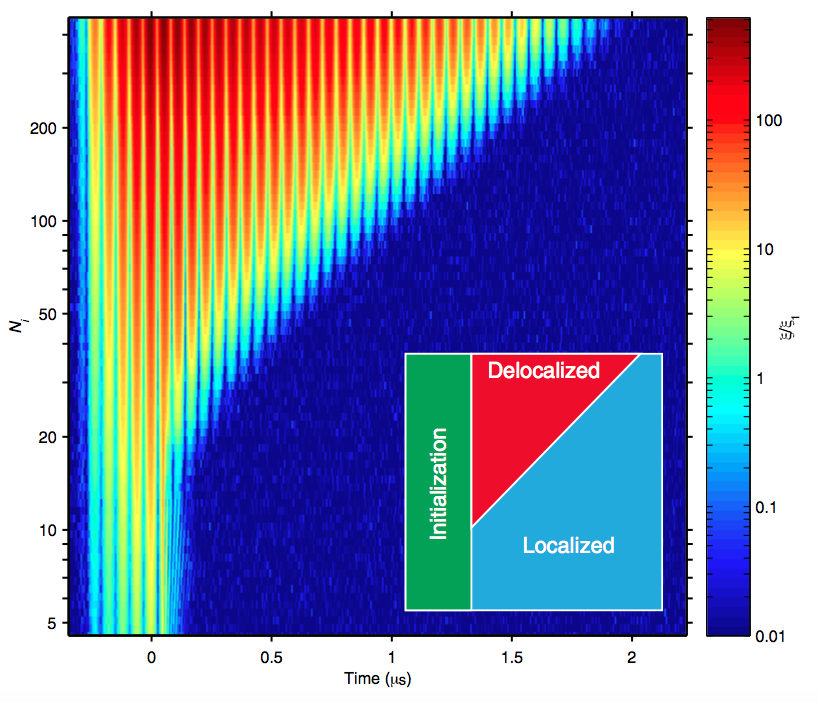}
\caption{
Phase diagram of the dissipation-induced delocalisation-localisation transition for the two-site system. Homodyne signal from the initially unoccupied site as a function of initial photon number ($N_i$) and time. For large enough $N_i$ the system starts in the delocalised regime, but as time progresses and photons are lost from the cavity, the system enters the localised regime. Taken from Raftery et al.~\cite{RafteryHouck2014}.
}
\label{fig:selftrapping}
\end{figure}

Such dissipation-induced transition has been observed experimentally in a superconducting circuit platform by Rafery et al.~\cite{RafteryHouck2014}. In this experiment, the atom-photon interaction is initially turned off by detuning the superconducting qubits out of resonance, and the resulting linear system is driven by a coherent pulse. Coherent oscillations of photons automatically prepare the required initial state with imbalance 1 at a certain time, at which point the interaction is ramped up by bringing the qubits into resonance. Figure \ref{fig:selftrapping} depicts the experimentally measured phase diagram, in which the homodyne signal of the photons from the initially unoccupied site is shown as a function of the initial mean photon number $N_i$ and time. Up to about $N_i \approx 20$, the atom-photon interaction strength $g$ is greater than the critical strength $g_c^{qu}$ and the photons are localised. Increasing the initial mean photon number beyond $N_i > 20$, $g$ becomes smaller than $g_c^{qu}$ and the photons start to oscillate between the two sites. However, as the system loses photons, the localised regime is recovered as the total number of photons in the system dips below some critical number.

\subsection{Fermionisation of hardcore photons}
In Sect.~\ref{sect:superbunching}, we have seen that the local second-order intensity correlation function reveals anti-bunching behaviour in the strongly interacting limit ($U/J \gg 1$). The physical picture is that two-photons cannot occupy the same site because of strong repulsion between them. This is reminiscent of the fermionic behaviour, where the Pauli exclusion principle prohibits two fermions from occupying the same state. In fact,
in 1 dimensional systems in equilibrium, strong interactions between bosons are known to induce fermionic behaviour \cite{Cazalilla}. As the on-site nonlinearity $U$ in the BH model approaches infinity, double occupancy of any resonator is completely suppressed and the bosonic wave function of the system can be written in terms of the wave function of free fermions \cite{Girardeau}.  That such fermionisation can be observed in driven dissipative CRAs was first noticed by Carusotto et al.~\cite{Carusotto}. By calculating the spectrum and intensity-intensity correlation functions of the output light in the NESS, it was shown that the strongly correlated many-body nature of the photons can be readily observed. Here we follow the work by Grujic et al.~\cite{GrujicAngelakis2012}, which investigated an analogous phenomenon in the JCH system. Readers are referfed to Ch.~[Gerace] or the original article \cite{Carusotto} for further details on the BH system. 

Consider a homogeneously driven 3-site cyclic array of nonlinear resonators. The eigenstates of this system in the hardcore regime can be determined by the aforementioned bose-fermi mapping. These eigenstates are conveniently labelled by quasi-momenta (of the Bloch modes), where the lowest eigenstates are found to be the one-particle state at zero momentum $|q=0\rangle$, and the two-particle state $|q,-q\rangle$ with equal and opposite momenta $q=\pi/3$ \cite{Carusotto}. 

\begin{figure}[ht]
\centering
\includegraphics[width=8cm]{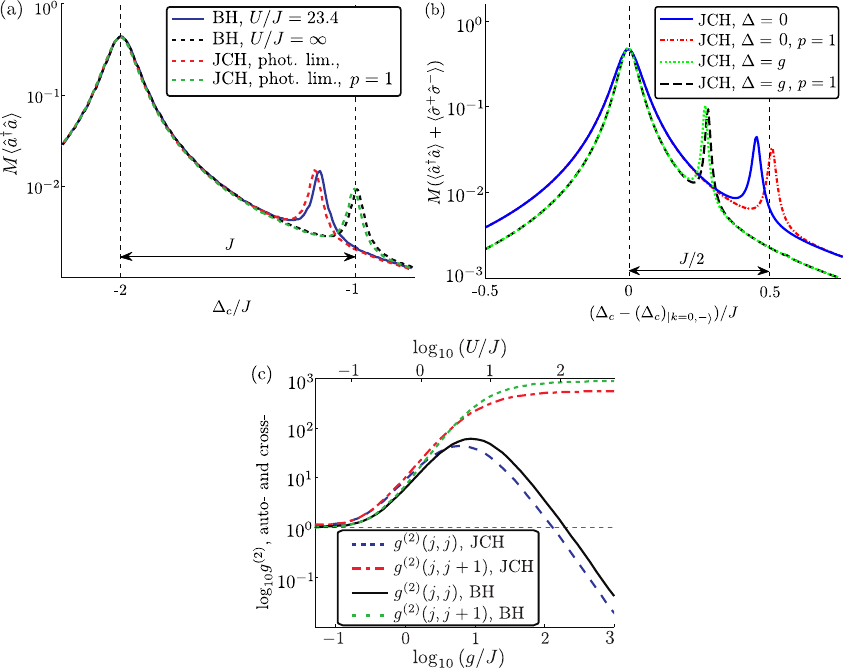}
\caption{
(a) Steady state photon number as a function of laser-cavity detuning for both the BH and JCH systems. One particle peaks are located on the left ($\Delta_c = -2J$) and two-particle peaks on the right. Strongly nonlinear cases are shown by the solid blue (BH) and the dotted red (JCH) curves, whereas the strictly hard-core limit is shown by the dotted black (BH) and the dotted green (JCH) curves. Parameters are $M=3$ cavities, $J/\gamma = 20$, $\Omega/\gamma = 0.5$, $\Delta/g = -10$, and $g/\gamma \approx 2 \times 10^5$. (b) Analogous photon number spectra for the JCH model with a smaller atom-photon detuning $\Delta$. $g/\gamma \approx 800$ as calculated from Eq.~(\ref{ueff}). (c) The auto- and cross-correlations measured at the two-particle peak as a function of nonlinearity in both the BH and JCH ($\Delta = 2J$) models. Values of $U$ are determined from Eq.~(\ref{ueff}). Taken from Grujic et al.~\cite{GrujicAngelakis2012}.}
\label{fig:fermionisation}
\end{figure}

The first evidence of the presence of fermionised photons can be observed in the steady-state excitation number as shown in Fig.~\ref{fig:fermionisation}(a). For the BH model, we clearly see the two lowest-lying fermionic energy eigenstates $|q=0\rangle$ and $|\pi/3,-\pi/3\rangle$ at $\Delta_c = -2J, -J$ in the hard-core regime (black dashed curve). Also shown is an effect of finite $U/J$: the peak is red-shifted towards lower energy (black solid curve). The JCH system shows the almost identical behaviour in the `photonic regime' of polaritons, where the atom-photon detuning is large compared to the coupling strength $g$ such that the photonic contribution to polaritons dominates over the atomic contribution. Photon numbers in the hardcore regime of the JCH model are calculated by truncating to at most one polariton excitation in each cavity.

From these calculations we conclude that the (JC) polaritons are also fermionised in the strongly interacting regime. However,
achieving strongly interacting regime for the photonic polaritons requires a very large value of coupling strength ($g/\gamma \approx 2\times10^5$ has been used in Fig.~\ref{fig:fermionisation}(a)) to compensate for the large atom-photon detuning. So, the question arises: Can we observe the same phenomenon for a more modest value of $g$? Figure \ref{fig:fermionisation}(b) displays the total excitation number for $g/\gamma \approx 800$, showing that similar physics can be observed by reducing the values of $\Delta$ and $g$. It is interesting to note that the quantitative agreement to the fermionised limit is better achieved for $\Delta = g$.

Further evidence of fermionisation can be obtained by inspecting the second-order intensity correlations between photons. For this purpose, the auto-correlations $g^{(2)}_{j,j}$ and the cross-correlations $g^{(2)}_{j,j+1}$ are depicted in Fig.~\ref{fig:fermionisation}(c). The driving field is tuned at the two-particle resonance. We have already seen the behaviour of the auto-correlations in Sect.~\ref{sect:superbunching}. In the weakly nonlinear regime, the resonance peaks corresponding to different values of $N$ (the total number of photons in the array) overlap, and the photons inherit the Poissonian nature of the pump field \cite{Carusotto}. With increasing nonlinearity, the two-photon peak initially gives rise to strong photon bunching, before fermionisation takes place at which point the photons become strongly anti-bunched. Fermionisation of photons is further corroborated by the cross-correlations, which increases with the nonlinearity, meaning that the photons are preferentially located next to each other.

\subsection{Polariton crystallisation}

In the above scenario we assumed homogeneous driving, i.e., all sites were driven with the driving field of the same intensity and phase. Let us now see an example of what happens when the system is driven inhomogeneously. The case where the nearest-neighbour driving fields have a phase difference of $\pi/2$ ($\Omega_k = \Omega \exp(ik\pi/2)$) has been studied by Hartmann \cite{Hartmann10}. This system exhibits an intriguing phenomenon of photon crystallisation, which will be briefly reviewed here. We follow the presentation in \cite{GrujicAngelakis2012}, which generalised the BH system in the original proposal to the JCH system. Interested readers are recommended to read the original paper \cite{Hartmann10} for an in-depth analysis of the effects of the inhomogeneous driving.

\begin{figure}[ht]
\centering
\includegraphics[width=8cm]{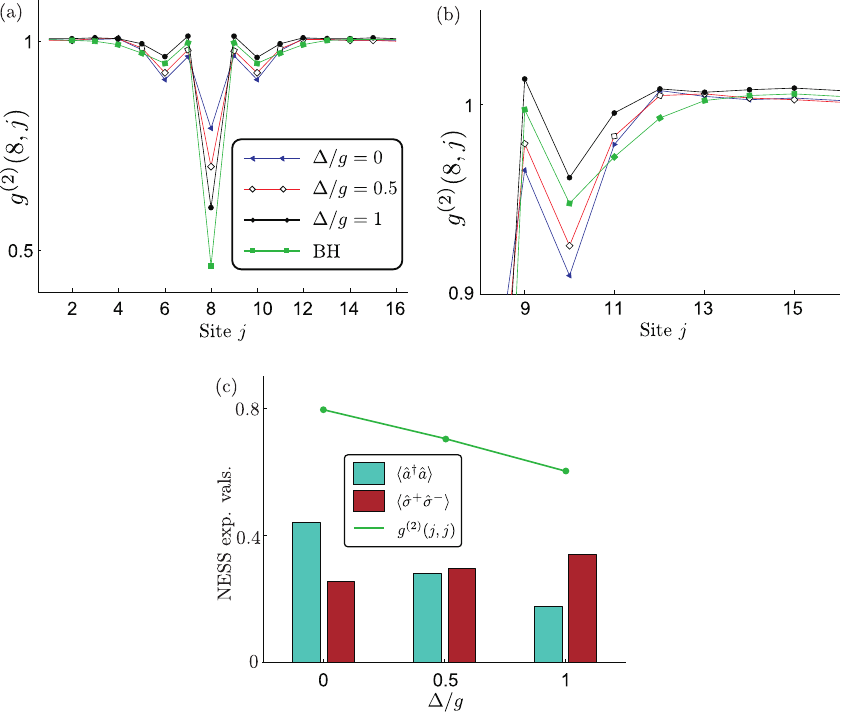}
\caption
{
(a) Steady state intensity correlations for a cyclic 16-site JCH system, driven by lasers exciting the $\pi/2$ momentum mode. Connecting lines are drawn to guide the eye. Parameters are $J/\gamma = 2$, $g/\gamma = 10$, and $|\Omega_k|/\gamma = 2$. Also shown (solid green) are the intenstiy correlations for the BH system with a `matched' Kerr nonlinearity $U/\gamma \approx 6$. (b) Closer view of the correlation function. (c) The atomic and photonic population in each resonator in the steady state for three different values of atom-resonator detuning, as well as the on-site intensity correlations. Taken from Grujic et al.~\cite{GrujicAngelakis2012}.}
\label{fig:crystallisation}
\end{figure}
The system consists of a cyclic ring of 16 cavities, with the above mentioned driving field arrangement. The latter creates a flow of polaritons around the ring, with momentum $\pi/2$, when the laser frequency is set to the lowest energy transition. The interplay between the flow and on-site interaction gives rise to the phenomenon of polariton crystallisation: there is a larger probability of finding photons in neighbouring cavities than in cavities further apart. This is evident in the second-order correlation function as shown in Fig.~\ref{fig:crystallisation}(a) and (b): the on-site anti-bunching is accompanied by nearest-neighbour intensity correlations that are stronger than correlations between more distant cavities. The interpretation is that the polaritons form dimers across two neighbouring sites which then move together around the array.

Both the BH and JCH systems clearly yield similar qualitative results, although there are qualitative differences in the small detuning limit considered here. As noted in the previous subsection, only in the `photonic regime'--where the photon-atom detuning is very large and g even larger--would one expect a quantitative match. Furthermore, the atom-photon detuning provides an additional knob in the JCH system that is absent in the BH system: it controls the composition of the polaritons. The photonic and atomic contributions to the polaritons are depicted in Fig.~\ref{fig:crystallisation}(c) for three different values of atom-cavity detuning. In this regime, the atomic contribution increases with $\Delta$, which explains the enhanced on-site anti-bunching with increasing $\Delta$.

\subsection{Phases in 2D arrays}

So far, we have limited our attention to 1D arrays, for which the steady-state or time-evolution could be studied numerically using either exact diagonalisation methods or the MPS formalism. Now we would like to discuss the physics of 2D arrays, where a whole new range of physics are waiting to be explored. However, there is a slight problem. In 2D, the application of tensor network ansatz does not work very well due to an unfavourable scaling of entanglement and are therefore difficult to investigate numerically. For this reason studies up to now have either focused on small systems or employed variants of mean-field analysis. Here, we give a brief survey of recent studies of 2D CRAs based on the latter.

Initially, the dynamics of interacting photons after an initial preparation in the Mott state were studied by Tomadin and coworkers \cite{TomadinImamoglu2010}. To solve the master equation, the authors employed the cluster mean-field approach, which reduces the whole system to a cluster of cavities plus mean-fields representing the rest of the system it is interacting with. For a single-site-cluster case, this amounts to making the mixed state equivalent of the usual mean-field approximation $\hat{a}_i^\dagger \hat{a}_j \rightarrow \langle \hat{a}^\dagger_i\rangle \hat{a}_j + \hat{a}^\dagger_i \langle \hat{a}_j \rangle$. In the typical quench scenario, the dynamics of interacting photons show characteristic differences depending on whether the system parameters are in the Mott or superfluid regime. Such signatures are also observed in the dissipative cases and are manifest in the coherence of the cavity emission. In the Mott-regime the equal-time second-order intensity correlation function stays zero, while in the superfluid-regime it gains a finite value within the photonic lifetime.

A mean-field phase diagram of the steady-state density matrix in the driven dissipative scenario has also been explored. Firstly, the phase diagram of the driven dissipative 2D BH array was mapped out by Le Boit\'e et al.~\cite{LeBoite,LeBoite2}, who found regions of unstable, mono-stable, and bi-stable phases. Despite Mott-lobe-like structures characterised by one stable solution with photon-antibunching, the phase diagram is very different from its equilibrium counterpart, showing rich many-body physics arising in the non-equilibrium scenario.
Secondly, Jin and co-workers have investigated the phases of a 2D BH array, but with an extra component of nearest-neighbour cross-Kerr interaction ($V \sum_{i,j} \hat{n}_i\hat{n}_j$) \cite{Jin,Jin2}. Due to this extra term, the NESS phases were shown to be classified into a uniform phase and a checkerboard phase. In the uniform phase, two neighbouring cavities have equal densities whereas in the checkerboard (or staggered) phase they have a finite density-difference. Furthermore, the NESS can show non-trivial oscillatory behaviour, hinting at a non-equilibrium analog of the supersolid phase.

Clearly, the validity of these mean-field results should be confirmed with approaches that take into account longer-ranged quantum correlations across the lattice. Numerical methods that go beyond the cluster mean-field approaches are under development. These include a hybrid real-space renormalisation group approach called `Corner space renormalisation method' \cite{Finazzi} and a self-consistent projection operator theory that derives an exact equation of motion for the reduced density matrices of an arbitrary sub-lattice \cite{Degenfeld-Schonburg}. 

\subsection{Quantum Hall physics with light}

It is possible to engineer artificial gauge fields for photons in a 2D CRA and thereby achieve photonic quantum Hall systems. There are several proposals to prepare and probe (integer or fractional) photonic quantum Hall states, but we will focus on recent developments on driven dissipative signatures in photonic quantum Hall systems and refer the interested reader to Ch.~[Hafezi].

In a lattice system, gauge fields such as magnetic fields can be included by the so-called Peierls substitution:
\begin{equation}
J \sum_{\langle i,j \rangle} \hat{a}^\dagger_i \hat{a}_j + {\rm h.c.} \rightarrow J \sum_{\langle i,j \rangle} e^{i\phi_{i,j}}\hat{a}^\dagger_i \hat{a}_j + {\rm h.c.}.
\end{equation}
Essentially, a particle traveling around any loop accumulates a phase factor corresponding to the sum of $\phi_{i,j}$ over the loop. It is possible to engineer such site-dependent hopping amplitudes in two-dimensional CRAs, to create an artificial magnetic field. Then, in analogy to the electronic systems, the presence of the on-site interaction gives rise to a photonic fractional Hall effect.
The latter can be checked by performing measurements on the output photons and calculating the overlap with the bosonic Laughlin state \cite{Umucalilar}, or can be inferred from the second-order intensity correlation function of the collective mode \cite{Hafezi}. Here, we we review the work by Umucal{\i}lar and Carusotto \cite{Umucalilar}. 

\begin{figure}[ht]
\centering
\includegraphics[width=8cm]{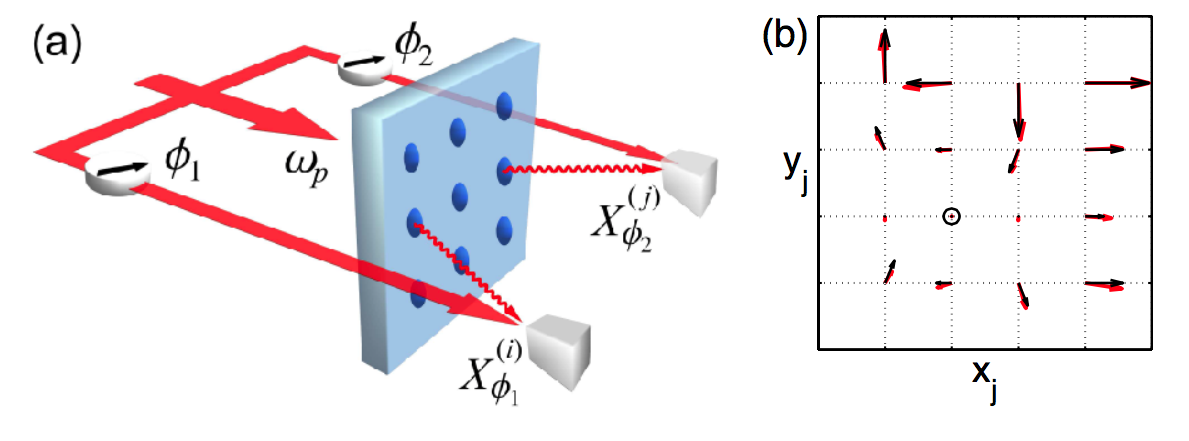}
\caption{
(a) Schematic illustration of the experimental set-up to measure the two-photon amplitude $\psi_{ij}$. (b) Comparison of the normalised two-photon amplitude for a fixed $i=i_0$ (red thick arrows) and the generalised Laughlin function (black thin arrows). The reference site $i_0$ is marked by a circle. Reprinted with permission from Umucal\i lar and Carusotto \cite{Umucalilar}. Copyright (2012) by the American Physical Society.
}
\label{fig:fractionalHall}
\end{figure}

Consider a 4 by 4 lattice in the hard-core limit $U/J \rightarrow \infty$, which is weakly pumped such that only up to 2 photons are allowed in the system at any given time. The effective magnetic field strength $B$ is set such that the number of flux quanta, the total magnetic flux divided by the flux quantum, is 4. This choice gives the filling fraction 1/2 when there are two photons in the system. In a conventional fractional quantum Hall system, the ground state is described by the bosonic analogue of a generalised Laughlin wave function \cite{Haldane}. However, the steady-state of the proposed driven dissipative set-up surely does not coincide with this state because there are 0 and 1 photon manifolds. Interestingly, though, the two-photon sub-manifold of the steady-state density matrix exhibits a strong overlap with the generalised Laughlin wave function. This has been shown in \cite{Umucalilar}, where the two-photon amplitude $\psi_{ij} \equiv {\rm Tr}[\rho_{ss}  \hat{a}_i \hat{a}_j]$ has been calculated numerically and  compared with the bosonic Laughlin state. Note that $\psi_{ij} $ can be measured experimentally by performing multiple homodyne measurements as schematically illustrated in Fig.~\ref{fig:fractionalHall} (a). An example of a numerically calculated $\psi_{ij}$ is shown in Fig.~\ref{fig:fractionalHall} (b), along with the corresponding Laughlin wave function. The system is driven at a two-photon resonance and weak dissipation $\gamma/J = 0.03$ and weak driving $\Omega/\gamma = 0.1$ are assumed. An overlap of up to $98.9 \%$ can be achieved for $\gamma/J = 0.002$, while for a more realistic loss rate $\gamma/J = 0.05$, the overlap value is $90.0 \%$. These high overlap values show that despite the losses and driving, one can prepare strongly-correlated photonic states which are analogues of the Laughlin state of fractional quantum Hall systems.

A driven dissipative analogue of the integer quantum Hall effect has also been proposed in a linear 2D CRA, where an experimentally viable scheme to measure the global Chern number (and even the local Berry curvature) was devised \cite{Ozawa}. In this proposal, a photonic analogue of the integer quantum Hall system is investigated, in which there is a constant force driving a current. Such a force can be modelled by a position dependent energy term $H_{f} = F\sum_{m,n} n \hat{a}^\dagger_{m,n} \hat{a}_{m,n}$ (force in -y direction). Recall that in integer quantum Hall systems the driven current induces a Hall current in the x-direction, whose conductivity is quantised. This quantisation is robust against various imperfections due to its topological nature, and is captured by the Chern number associated with each Bloch band. Remarkably, it is possible to measure the Chern number in a driven dissipative set-up as proposed by Ozawa and Carusotto \cite{Ozawa}. Alternative methods, based on tuning twisted boundary conditions, to experimentally measure the topological invariants in driven-dissipative settings were proposed by Hafezi \cite{Hafezi2014} and Bardyn and coworkers \cite{Bardyn2014}. 

\begin{figure}[ht]
\centering
\includegraphics[width=6cm]{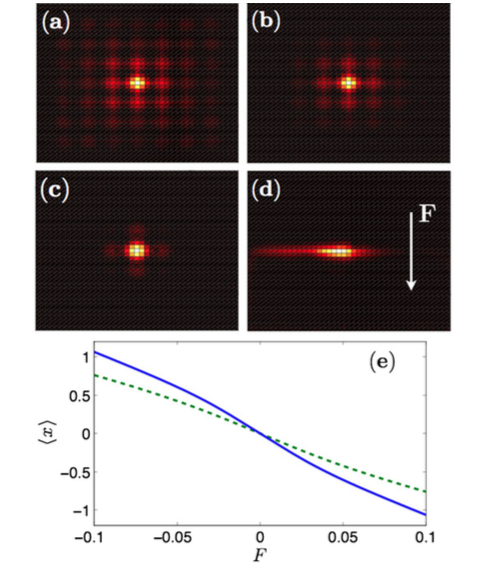}
\caption{
(a)--(d) Distributions of the photon amplitudes $|a_{m,n}|$ on a 41 $\times$ 41 lattice. (a) $F=0$, $\gamma = 0.01J$, and a pump frequency within the lowest band. (b) The same as in (a) but with larger loss rate $\gamma = 0.02J$. (c) The same as in (a) but the pump frequency lies within a band gap. (d) The same as in (a) but with $F = 0.1J$. (e) Displacement $\langle x \rangle$ as a function of $F/J$, for a pump frequency within the lowest band. The solid blue line is for $\gamma/J = 0.05$ and the dashed green line is for $\gamma/J = 0.08$. Reprinted with permission from Ozawa and Carusotto \cite{Ozawa}. Copyright (2014) by the American Physical Society.
}
\label{ozawa}
\end{figure}

To understand the method by Ozawa and Carusotto, consider a linear 2D array (of size 41 $\times$ 41), with the central site driven by a coherent field of amplitude $\Omega$. Then, because the system is linear, we can replace the operators $\hat{a}_{m,n}$ by complex amplitudes $a_{m,n}e^{-i\omega_0t}$. When the driving force is absent, i.e., $F=0$, the injected photons either disperse through the lattice (Fig.~\ref{ozawa}(a) and (b)) or stay localised within the central site  (Fig.~\ref{ozawa}(c)), depending on whether the driving frequency lies within a Bloch band or a band gap. With the driving force turned on ($F=0.1J$), however, the photons clearly `travel' towards the left as shown in Fig.~\ref{ozawa}(d). It is possible to show that the displacement in the center of mass $\langle x\rangle \equiv \sum_{m,n} m|a_{m,n}|^2/\sum_{m,n} |a_{m,n}|^2$ is related to the Berry curvature $\Omega_{B}(k)$ by the formula \cite{Ozawa}
\begin{equation}
\langle x\rangle = F\frac{\int_{\rm MBZ} \gamma \Omega_B(k)n^2(k)}{\int_{\rm MBZ} n(k)}.
\end{equation}
Here, MBZ stands for the magnetic Brillouin zone; $n(k) = [(\omega_L - E(k))^2 + \gamma^2]^{-1}$, where $\omega_L$ is the driving frequency; $E(k)$ is the energy dispersion of the corresponding band. In the large-loss limit where $\gamma$ is bigger than the width of the Bloch band under consideration but smaller than the band gap, the above formula reduces to
\begin{equation}
\langle x\rangle \approx \frac{qCF}{2\pi\gamma},
\end{equation}
where $C$ is the Chern number. On the other hand, if the loss rate is very small compared to the width of the Bloch band, the formula reduces to
\begin{equation}
\langle x \rangle \approx \frac{\bar{\Omega}_B(\omega_L)F}{2\gamma},
\end{equation}
where $\bar{\Omega}_B(\omega_L)$ is the averaged Berry curvature on the $E(k)=\omega_L$ curve. Examples of the linear dependence of $\langle x \rangle$ on $F$ is shown in Fig.~\ref{ozawa}(e).

Therefore, depending on the value of the loss rate, one can either directly measure the Chern number or the local Berry curvature. For the lowest Bloch band, which has $C=-1$, the theoretically calculated value using the above method was shown to be as close as -0.97, demonstrating that the signature of the integer quantum Hall effect can be directly measured in the driven dissipative set-up. 
Furthermore, the ability to measure the Berry curvature allows one to detect nontrivial properties of the photonic honeycomb lattice in the absence of an external magnetic field. In the latter, there are two bands near the Dirac points whose degeneracy is lifted when two neighbouring sites have different energies. Even though the total Chern number vanishes in the absence of a synthetic magnetic field, the two bands are known to possess nontrivial Berry curvatures. Because the Chern number vanishes, one needs to resolve the Berry curvature in order to detect the system's nontrivial properties, which can be done using the above method.

\subsection{Other works}

There are a number of interesting works on driven dissipative CRAs that we could not cover due to spatial limitations. Here we provide a brief survey of some of these, so that the interested readers can refer to original articles. 

First of all, there are many works that investigate properties of photons emitted from a dimer. Liew and Savona discovered the phenomena of `unconventional photon blockade', where photon blockade is achieved in the weakly Kerr-nonlinear regime in which anti-bunching would otherwise be absent \cite{Liew}. This was shown to result from an interference of excitation pathways \cite{Bamba} and was generalised to a coupled resonator with second-order nonlinearity \cite{GeraceSavona2014}. The fluorescence spectrum and the spectrum of the second-order intensity correlation function of the JC dimer were calculated in \cite{Knap}, both for coherent and incoherent driving fields, while photon statistics in a BH dimer were investigated with emphasis on photon anti-bunching in \cite{Leib, FerrettiGerace2010}. Scaling of this anti-bunching behaviour with increasing array size was studied within a mean-field theory \cite{Nissen}. In a slightly different setting, transport of photons in a BH array of up to 60 sites was studied using an MPS method adapted to mixed states \cite{Biella}. Here, only the first cavity is driven and the underlying many-body states were shown to be visible in the transmission spectrum. Transport of quantum light in a BH dimer was also studied, showing that more detailed information about the underlying correlated states is revealed when one uses two photons of different energies instead of a coherent field as the input \cite{Lee}. 

Other works look at connections with well-known condensed matter systems. A quantum optical analogue of the Josephson interferometer in a three-site set-up was proposed by Gerace and co-workers \cite{GeraceFazio2009}. There, in analogy to the Josephson junction, two end cavities are driven by coherent fields with different phases, while the middle cavity contains single-photon nonlinearity.  The interplay between tunneling and interaction gives rise to rich physics in the steady-state, details of which can be found in Ch.~[Gerace]. A driven dissipative realisation of the Kitaev chain \cite{Kitaev} was proposed by Bardyn and \.{I}mamo\u{g}lu \cite{Bardyn}, showing that Majorana zero modes can be created and detected in a CRA. To obtain the so-called $p$-wave pairing term, the authors propose to use parametric driving in the strongly-interacting regime. Photons are then effectively spin-1/2 particles, allowing one to use Jordan-Wigner transformation to achieve the Kitaev chain. 

\section{Summary and outlook}
In this chapter, we have reviewed out-of-equilibrium many-body physics in coupled resonator arrays. A short survey of possible experimental platforms was provided, along with two theoretical models (Jaynes-Cummings-Hubbard and Bose-Hubbard) to describe them as well as the master equation description of driven dissipative CRAs. 
Then we reviewed a range of many-body non-equilibrium phenomena, starting from those arising in a two-site set-up and finishing with phenomena in  2D arrays. A brief survey of many interesting works that could not be covered in detail was then given. 

The study of driven dissipative CRAs has only recently begun and there are many exciting avenues to be explored. Experimentally, it is still quite challenging to build an array of resonators with large nonlinearity, although splendid progress is being made in the field of circuit-QED as we have seen in Sect.~\ref{sect:frozen}. Controlling the fluctuations of local parameters is another limitation. Theoretically, one can look at ways to enhance the effects of nonlinearity, as in unconventional quantum blockade, to help ease the requirement of strong nonlinearity. More pressing is the development of theoretical tools to study driven dissipative 2D arrays. As we have alluded to earlier, 1D systems can be simulated efficiently using the MPS formalism, but 2D system are waiting for further developments. There is already progress in this direction as we briefly mentioned at the end of the last section, and more will surely follow. With the on-going developments in both the experimental and theoretical fronts, many more exciting discoveries will surely follow in this nascent field.
\begin{acknowledgement}
C. Noh and D. G. Angelakis would like to acknowledge the financial support provided by the National Research Foundation and Ministry of Education Singapore (partly through the Tier 3 Grant ``Random numbers from quantum processes''), and travel support by the EU IP-SIQS. S. R. Clark and D. Jaksch acknowledge support from the European Research Council under the European Union's Seventh Framework Programme (FP7/2007--2013)/ERC Grant Agreement no.~319286 Q-MAC.
\end{acknowledgement}

\section*{Appendix: Matrix Product States and quantum trajectories}
In this Appendix we provide some more details about MPS methods and quantum trajectories approach. First we specify the problem to be solved in general.

\subsubsection*{Quantum master equation}
It is known that the most general form of a Markovian time-local master equation that preserves the trace, Hermiticity, and positivity of the system's density matrix $\rho$ must be of so-called Lindblad form \cite{carmichael,breuer_petruccione}: 
\begin{equation}
\frac{d \rho}{dt} = -i [\hat{H}, \rho] + \sum_{\alpha} \left ( \hat{L}_\alpha \rho \hat{L}_\alpha^\dag  - \frac{1}{2} \left \{ \hat{L}_\alpha^\dag \hat{L}_\alpha, \rho \right \}\right ), 
\label{eq:theory_master_equation_normalised_lindblad_form}
\end{equation}
where the Lindblad operators $\hat{L}_\alpha = \sqrt{\gamma_\alpha} A_\alpha$ and $\hat{H}$ is the Hamiltonian describing the unitary part of the system's dynamics. The $\gamma_\alpha$ are identified as the characteristic rates at which `jump operators' $A_\alpha$ act on the system. In the case of CRAs the jump operators are the bosonic mode operators $\hat{a}_j$ or atomic ladder operator $\hat{\sigma}^-_j$ which act locally on a site $j$. The locality of the jump operators and the nearest-neighbour nature of the JCH and BH Hamiltonians in 1D are crucial to the applicability of MPS methods to solve this problem. 

\subsubsection*{Matrix Product States}
Generally we consider an $L$ site chain composed of subsystems with $d$-dimensional Hilbert space spanned by states enumerated as $|j\rangle$, with $j \in \{1, \cdots, d \}$. The Hilbert space for the total system is then spanned by the product basis $\ket{{\bf j}} = |j_1, \cdots, j_L \rangle \equiv | j_1 \rangle \otimes \cdots \otimes |j_L \rangle$. An arbitrary pure state of the system is then
\begin{eqnarray}
|\Psi \rangle & = & \sum_{\bf j} c_{j_1j_2\cdots j_L} \ket{\bf j}, \label{eq:general_lattice_expansion_psi}
\end{eqnarray}
which is described by $d^L$ complex amplitudes $c_{j_1j_2 \cdots j_L}$. To circumvent this intractable description the MPS ansatz parameterises the amplitudes of a state in terms of a product of matrices indexed by the local configurations as
\begin{equation}
|\Psi \rangle = \sum_{\bf j} \left(\bar{L}^{\rm T}\mathbf{A}^{j_1} \mathbf{A}^{j_2} \cdots \mathbf{A}^{j_L}\bar{R}\right) \ket{\bf j}.
\label{eq:MPS_parameterisation}
\end{equation}
Here the $\mathbf{A}^{j_m}$ are a set of $d$ matrices (one for each local state $\ket{j_m}$) of dimension $\chi \times \chi$, while $\bar{L}$ and $\bar{R}$ are boundary vectors which collapse the product into a scalar complex amplitude. This ansatz has only $Ld\chi^2$ complex numbers, so if $\chi$ is fixed and small it provides a compact class of states \cite{schollwock_ann_phys}. 

A special feature of MPS is that the dimension of the matrices $\chi$ is directly related to the entanglement of the system when it is bipartitioned. If a state is weakly entangled then it will possess an accurate MPS representation with a small fixed $\chi$. Given an MPS, reduced density matrices for subsets of sites, norms and expectation values of observables can be efficiently computed, essentially via products of the state's constituent matrices \cite{perez_garcia,schollwock_ann_phys}. These features become especially useful for 1D systems which are described by a Hamiltonian $\hat{H} = \sum_m \hat{h}_{m,m+1}$ composed of at most nearest-neighbour terms. Many such systems are now known to possess weakly entangled ground states and low-lying excited states. Thus an MPS description of them is both natural and effective \cite{schollwock_ann_phys}. 

To find ground states and excited states of $\hat{H}$ the highly successful density matrix renormalisation group (DMRG) method \cite{schollwock_ann_phys,white_a,white_b,schollwock_rmp} was developed which is essentially a variational minimiser over the MPS ansatz. For our purposes here however the extension of DMRG to time-evolution via the time evolving block decimation (TEBD) algorithm is most relevant \cite{vidal_a,vidal_b}. Here we sketch some of the key points to this method. Our task is to apply unitary time-evolution described by the propagator $\hat{U}(\delta t) = \exp(-i\hat{H}\delta t)$ to an initial MPS for a small time step $\delta t$. 

First, we break up the exponential using a Suzuki-Trotter decomposition \cite{suzuki} such as
\begin{equation}
\hat{U}(\delta t) = \Bigg[\prod_{{\rm odd}~m} \exp(-i\hat{h}_{m,m+1}\delta t) \Bigg] \Bigg[\prod_{{\rm even}~m} \exp(-i\hat{h}_{m,m+1}\delta t)\Bigg] + \mathcal{O}(\delta t^2),
\end{equation}
where we have utilised that all the odd Hamiltonian terms $\hat{h}_{m,m+1}$ commute among themselves, and similarly for the even terms. Higher order Trotter decompositions have a similar form. What we are left with is a product of unitary operators (or gates) $\exp(-i\hat{h}_{m,m+1}\delta t)$ acting on nearest-neighbouring pairs of sites which have to be applied to the MPS initial state in sequence. 

Second, upon applying a gate $\exp(-i\hat{h}_{m,m+1}\delta t)$ to an MPS the matrices ${\bf A}^{j_m}$ and ${\bf A}^{j_{m+1}}$ for the two sites $m$ and $m+1$ get merged together into a joint matrix $\mathbf{B}^{j_m,j_{m+1}}$. To bring the state back into MPS form the matrix $\mathbf{B}^{j_m,j_{m+1}}$ needs to be factorised. This is achieved by using a singular value decomposition (SVD) which breaks $\mathbf{B}^{j_m,j_{m+1}} \mapsto \tilde{{\bf A}}^{j_m}\tilde{{\bf A}}^{j_{m+1}}$ yielding new updated matrices describing the evolved MPS \cite{vidal_a,vidal_b}. During this operation the dimensions of the new $\tilde{\bf A}$ matrices can grow and so after repeated applications of gates the MPS can eventually become intractably large. However, the singular values outputted by the SVD provide a quantitative means of truncating the dimension down, essentially compressing the state and thus ensuring that the MPS remains tractable. This truncation will only be accurate if the evolution does not generate too much entanglement. Once all the gates in the decomposition are applied we have an MPS approximation for $\ket{\Psi(\delta t)} = \hat{U}(\delta t)\ket{\Psi}$, and the procedure can be repeated to evolve further in time \cite{schollwock_ann_phys}.

\subsubsection*{Quantum Trajectory Algorithm} \label{sec:traj_algorithm}
Rather than simulating the dynamics of an open system by evolving its density matrix $\rho$ directly, an alternative approach is to propagate stochastic realisations of individual system state vectors $\{ | \Psi_i(t) \rangle \}$ \cite{gardiner_parkins_zoller,dum_gardiner,dalibard_castin_molmer}. These `quantum trajectories' are piecewise deterministic processes, interrupted randomly by `quantum jumps' due to the system's interaction with the environment. An approximation to the density matrix may be obtained by averaging the contributions $\Pi_i = \{ | \Psi_i(t) \rangle \langle \Psi_i(t) | \}$ of many trajectories \cite{plenio_knight}. The details of the `unravelling' of the Lindblad master equation into stochastic wave functions are described in Ref. \cite{breuer_petruccione}. Here we give a practical recipe on how to generate quantum trajectories.

For a set of Lindblad operators $\{ \hat{L}_\alpha\}$, we construct a non-unitary effective Hamiltonian $\hat{H}_{\rm eff}$:
\begin{equation}
\hat{H}_{\rm eff} = \hat{H} - \frac{i}{2} \sum_\alpha \hat{L}^\dag_\alpha \hat{L}_\alpha, 
\label{eq:effective_ham_defn}. 
\end{equation}
Evolution under the Schr\"odinger equation with this effective Hamiltonian leads to a decay in the norm of the wave-function as a consequence of interactions with the environment. Trajectories are then generated as follows. 

\begin{enumerate}
\item Begin the simulation by initialising the system in a state $| \Psi \rangle = | \Psi(0) \rangle$. 
\item Draw a randomly chosen number $r \in [0,1]$. 
\item Evolve the system under the action of the non-Hermitian Hamiltonian of Eq.~(\ref{eq:effective_ham_defn}). This will involve computing $\ket{\Psi(t)} = \exp(-i\hat{H}_{\rm eff} t)\ket{\Psi(0)}$. Due to the decaying norm of the state, physical expectation values are calculated during this evolution using the normalised form of the state $|\Psi (t) \rangle / || | \Psi (t) \rangle ||$. 
\item Continue the evolution until the norm falls below the randomly chosen number, i.e. when the condition $|| |\Psi (t_{\rm jump}) \rangle || < r$ is first met. One of the possible quantum jumps now occurs. 
\item The jump operator to be applied is chosen by first generating the normalised probability distribution
\begin{equation}
P_\alpha = \frac{|| \hat{L}_\alpha \Psi(t_{\rm jump}) \rangle||^2 }{\sum_\beta || \hat{L}_\beta \Psi(t_{\rm. jump}) \rangle||^2 },
\label{eq:normalised_jump_dist}
\end{equation}
then randomly selecting a jump operator index $j$ from this weighted distribution. The normalised state after the application of the chosen operator is 
\begin{equation}
\frac{\hat{L}_j | \Psi (t_{\rm jump}) \rangle}{|| \hat{L}_j | \Psi (t_{\rm jump}) \rangle ||}. 
\end{equation}
\end{enumerate}
We continue looping over steps 2-5 until the end of our simulation window. Crucially all the steps in this algorithm, such as the time-evolution in step 3, the computation of norms in step 4 and the application of a jump operator in step 5, can be implemented efficiently on MPS so long as the Hamiltonian and Lindblad operators act locally or on nearest-neighbour sites.

\end{document}